\documentstyle[12pt,psfig,amsfonts]{article}

\newcommand{\NP}[1]{ Nucl.\ Phys.\ {#1}}

\newcommand{\PL}[1]{ Phys.\ Lett.\ { #1}}

\newcommand{\PR}[1]{Phys.\ Rev.\ { #1}}
\newcommand{\PRL}[1]{ Phys.\ Rev.\ Lett.\ { #1}}

\newcommand{\Od}{{\cal O}}

\newcommand{\im}{\mbox{Im}\,}
\newcommand{\re}{\mbox{Re}\,}

\newcommand{\sv}{\vert \vec{S} \vert}
\newcommand{\stsv}{(S_0,\vert \vec{S} \vert)}

\newcommand{\be}{\begin{equation}}
\newcommand{\ee}{\end{equation}}
\newcommand{\ba}{\begin{eqnarray}}
\newcommand{\ea}{\end{eqnarray}}

\newcommand{\ra}{\rangle}
\newcommand{\la}{\langle}

\newcommand{\IZ}{{\Bbb Z}}
\newcommand{\IR}{{\Bbb R}}

\newcommand{\gsim}{\raise.3ex\hbox{$>$\kern-.75em\lower1ex\hbox{$\sim$}}}
\newcommand{\lsim}{\raise.3ex\hbox{$<$\kern-.75em\lower1ex\hbox{$\sim$}}}

\begin{document}
\baselineskip=20pt

\begin{center}
  {\Large \bf Finite temperature pion vector form factors in\\ Chiral
  Perturbation Theory}

\vspace{.5cm} {\large A. G\'omez Nicola
, F. J. Llanes-Estrada and J. R. Pel\'aez.}

\emph{Departamentos  de F\'{\i}sica Te\'orica I y II.
  Univ. Complutense. 28040 Madrid. SPAIN.}

\end{center}


\vspace{-.5cm}
\noindent
\rule{\textwidth}{.1mm}
\begin{abstract}

We discuss  the thermal behaviour of the  pion vector form factors
and calculate them in one-loop  Chiral Perturbation Theory. The
perturbative result is used to analyze the $T$-dependent
electromagnetic pion charge radius, obtaining a rough estimate of
the deconfinement critical temperature. Imposing thermal
unitarity, we generate the $\rho$ resonance pole for the form
factor in the center of mass frame. The $\rho$ peak height in the
modulus of the form factor decreases for increasing temperature,
while its width increases  and the peak position is slightly
shifted downwards for $T\simeq$ 150 MeV.  These results point in
the direction suggested by many analysis of dilepton production
data
      in relativistic heavy ion collisions.
\vspace*{.2cm}

 PACS: 11.10.Wx, 12.39.Fe, 11.30.Rd, 25.75.-q., 12.38.Mh.
\end{abstract}
\vspace{-.5cm}
\rule{\textwidth}{.1mm}
\section{Introduction}

Ongoing experiments on relativistic heavy ion collisions
have attracted much attention over the past years. After the
expected Quark-Gluon-Plasma cools down and hadronic matter forms,
a correct description of the system involves QCD at temperatures
below the chiral phase transition. In this regime, the medium
constituents are predominantly light mesons, whose very low-energy
dynamics is described by Chiral Perturbation Theory (ChPT)
\cite{we79,gale84}, the low-energy effective theory of QCD  based
on spontaneous chiral symmetry breaking. ChPT is a low-energy
expansion performed in terms of $p^2/\Lambda_\chi^2$ with $p$ a
typical meson energy or temperature and $\Lambda_\chi \simeq$ 1
GeV. It has been successfully applied to light meson dynamics and
also to describe the low-$T$ pion gas \cite{gale87}. Recently, we
have calculated in this framework the $\pi\pi$ thermal scattering
amplitude \cite{glp02} and its unitarization \cite{dglp02} showing
that chiral symmetry and unitarity alone provide a reasonable
description of the thermal  $\rho$ and $\sigma$ resonances,
without
  including them as explicit degrees of freedom (see also \cite{coimbra}).

One salient observable in heavy ion collisions  is the dilepton
spectrum: $e^+ e^-$ pairs are direct probes of the plasma
evolution, since they do not interact from the production point to
the detector. The observed spectrum in recent experiments
performed at the SPS collider for invariant masses between 200 MeV
and 1 GeV
 differs significantly from vacuum hadronic emission
models,  showing a global enhancement for center of mass energies
between 0.2-1 GeV \cite{SPS}. In addition, the DLS collaboration
at the BEVALAC accelerator has reported results with lighter ions
and lower colliding energies per nucleon where the low-mass
enhancement is even bigger and defies theoretical explanations
\cite{DLS}.

The explanation of the dilepton SPS enhancement has been the
subject of intense theoretical work
\cite{li95,koso96,sb,rawa99,dileptonvarios}. This effect is
particularly visible near the $\rho$ mass, around which the
spectrum flattens and is compatible
   with a widening of the $\rho$ \cite{li95,sb,rawa99,ele01}.
   Ignoring baryon density effects, the main contribution to the dilepton spectrum
at low energy stems from the annihilation of two thermal pions via
the emission of a virtual photon
$\pi^+\pi^-\rightarrow\gamma^*\rightarrow e^+e^-$
\cite{kaj86,gaka} where the $\rho$ is produced as an intermediate
resonant state. Thus, the production rate for dileptons is
governed by the pion form factor  \cite{kaj86,gaka} whose
in-medium modification in the pion gas is key to the spectrum
\cite{li95,koso96,sb,soko96}. Prior calculations have relied on
model dependent input, the closest to our approach being the
finite temperature form factor analysis in \cite{soko96}, a chiral
model with resonances explicitly included as independent fields in
the Vector Meson Dominance (VMD) framework. Another approach
\cite{dom94} uses a sum rule calculation to extrapolate to low
energies from the perturbative results of QCD. In this work we
first present a model independent
 study of the finite temperature effects on the pion
form factor using ChPT. In particular, after a brief analysis of
the different thermal form factors in Section 2, we obtain the
one-loop ChPT thermal calculation, we study the temperature
evolution of the pion electromagnetic charge and radius, obtaining
a merely qualitative estimate of the critical temperature of
deconfinement, and we check thermal unitarity. Second, in Section
4, we implement unitarity to describe the effect of the thermal
variation of the $\rho$ resonance mass and width in the form
factor. In section 5 we summarize our main results. Comparing with
our previous works  \cite{glp02,dglp02}, we remark that, although
the phase of the form factor is related with the amplitudes via
unitarity, the modulus, which we calculate here in ChPT, is an
independent object needed to describe relevant physical quantities
such as the charge radius or the dilepton rate.

\section{Finite temperature vector form factors}
\label{sec:ftff}

At $T\neq 0$, all  physical quantities may depend on the fluid
four-velocity, so that, with the usual choice of the fluid rest
frame, Lorentz covariance is lost while spatial rotation
covariance is still preserved. Therefore, the most general
expression  for the timelike vector form factor of charged pion or
electromagnetic $\pi^+\pi^-\gamma$ vertex, is \footnote{ An
equivalent parametrization is used in \cite{soko96}, where
$F_s=F$, $G_s=G'/(p_0-q_0)$ and $F_t=F+G'(p_0+q_0)/(p_0-q_0)+
G/(p_0-q_0)$. }: \ba
    \la \pi^+(p)\pi^-(p')\vert V_0(0)\vert 0\ra &=&
    q_0 F_t(S_0,\sv,q_0)\nonumber\\
\la \pi^+(p)\pi^-(p')\vert V_k(0)\vert 0\ra &=&
    q_k F_s(S_0,\sv,q_0)+S_k
    q_0 G_s(S_0,\sv,q_0)
    \label{ffgenform}
\ea
with $V_\mu$ the electromagnetic current,  $S=p+p'$, $q=p-p'$ and
$F_t$, $F_s$, $G_s$  even functions in $q_0$ (charge conjugation
invariance). Note that the effect of Lorentz covariance breaking
is twofold: on the one hand, the time and spatial components of
the current may depend on the three different functions $F_t$,
$F_s$ and $G_s$ and, on the other hand, those functions may depend
on three independent rotationally invariant variables, instead of
just one invariant variable ($s=S^2$) as in the $T=0$ case. Terms
containing $\epsilon_{kij}S^i q^j$, allowed by rotation invariance
and $C$ are forbidden by parity.

Of course, gauge invariance further restricts the above
expressions, imposing relations between the functions $F_{t,s},G$.
When taking the divergence of the current, it must be also taken
into account that the in-medium pion dispersion relation may be
different from the vacuum one, i.e, the pole of the pion
propagator is at $p^2=m_\pi^2+g(p_0,\vert \vec{p} \vert;T)$ with
$g$ a complex function. Altogether, this relates the form factors
and the function $g$. However, we will
 content ourselves here with one-loop ChPT, which gives the real (and
positive) leading $\Od(p^4)$ to $g$, that depends on $T$ but not
on energy and momentum \cite{gale87}.
 That is, to $\Od(p^4)$ the two pions in (\ref{ffgenform})
 can be treated as free with a $T$-dependent mass shift.
  Therefore, to $\Od(p^4)$ we have $q\cdot S=0$, that combined with
gauge invariance $\la \pi\pi\vert \partial_\mu V^\mu\vert0\ra=0$
 in (\ref{ffgenform}) yields:

\be S_0\left( F_t^{(1)}-F_s^{(1)}\right)=\sv^2 G_s^{(1)}
\label{ginlo} \ee where the superscript $(1)$ means the NLO
(one-loop) form factors. Remember that the ($T$-independent)
leading order is just $F_t^{(0)}=F_s^{(0)}=1$, $G_s^{(0)}=0$.

As we will see below, in one-loop ChPT $G_s^{(1)}\neq 0$ and
therefore there are {\em two} independent thermal form factors
$F_t^{(1)}\neq F_s^{(1)}$. Moreover, there is an additional
simplification valid to one loop: the $F_t,F_s,G_s$ functions do
not depend on $q_0$. The gauge invariance condition (\ref{ginlo})
coincides with the analysis performed in \cite{soko96}.

\section{One loop ChPT calculation}
\label{sec:loopcal}

In this section we follow similar steps as in our pion scattering
calculation \cite{glp02}. We calculate the time-ordered product of
two pion and one current fields in the imaginary time formalism
(ITF) of Thermal Field Theory \cite{lebellac}. This we continue
analytically for continuous external pion energies
 and connect to the form factor through the  LSZ reduction
 formula. Such thermal amplitudes correspond to retarded real-time Green functions  and
 have the correct analytic and unitarity structure
 \cite{glp02,dglp02}, properties of special interest in this study.

\begin{figure}[h]
\hspace*{1.3cm} \hbox{\psfig{file=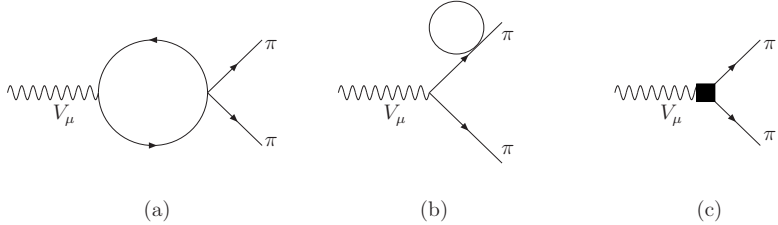,width=11cm}}
\caption{\rm \label{fig:diagrams} Diagrams contributing to the
pion form factors to one loop in ChPT.}
\end{figure}

 In the ITF,
 the lagrangian and electromagnetic current coincide with those in
ordinary $T=0$ ChPT. The thermal modifications arise upon
 replacing all zeroth momentum components by discrete
frequencies $k^0\rightarrow i\omega_n=2\pi i n T$ and the loop
integrals  by Matsubara sums, i.e., $ \int \! \frac{dk^0}{2\pi}
\rightarrow i T \sum_{-\infty}^{n=\infty}$.
We draw in Figure \ref{fig:diagrams}
the diagrams (also the same as for $T=0$) contributing to the form factors
at NLO within the $SU(2)$ chiral lagrangian \cite{gale84}.

Diagram (a) gives, among other contributions,
 the imaginary part needed for unitarity (see section \ref{sec:pu})
 while diagram (b) is proportional to the $T$-dependent tadpole which is
real and affects the electromagnetic vertex through wavefunction
renormalization. Their ITF contributions to the current
expectation value, before performing the analytic continuation are
given by: \ba \la \pi^+\pi^-\vert V_\mu\vert 0\ra_{(a)}
&\rightarrow& -\frac{2 q^\nu}{f_\pi^2} \
T\sum_{n=-\infty}^{n=+\infty} \int \frac{d^{D-1}k}{(2\pi)^{D-1}} \
\frac{k_\mu
k_\nu}{(k^2-m_\pi^2)((k-S)^2-m_\pi^2)}\nonumber\\
\la \pi^+\pi^-\vert V_\mu\vert 0\ra_{(b)} &\rightarrow&
\frac{q_\mu}{f_\pi^2} \ T\sum_{n=-\infty}^{n=+\infty} \int
\frac{d^{D-1}k}{(2\pi)^{D-1}} \ \frac{1}{k^2-m_\pi^2}
\label{jmuoneloop} \ea where $S_0,k_0\in 2\pi i T \IZ$  and $D$ is
the space-time dimension. As for diagram (c), the black box stands
for the $\Od(p^4)$ lagrangian tree level contribution, and
therefore $T$-independent, proportional to the low-energy constant
$l_6$ \cite{gale84}.

The integrals appearing in (\ref{jmuoneloop}) can be related to
the $T$-dependent one-loop integrals discussed in the Appendix of
\cite{glp02}. We adhere to  the notation and conventions of that
work. Using the formulae  there quoted and after analytic
continuation, the form factors can be written in terms of three
independent thermal integrals: $J_2, J_0$, that are energy and
momentum dependent, and the constant tadpole integral $F_\beta$
\footnote{We have used $J_1\stsv=\frac{S_0}{2}J_0\stsv$, that
holds for any $S_0,\sv$.}: \ba \Delta
F_t^{(1)}\stsv&=&\frac{s}{2f_\pi^2 \sv^2}\left[
4\Delta J_2\stsv-S_0^2\Delta J_0\stsv-2\Delta F_\beta\right] \nonumber\\
\Delta F_s^{(1)}\stsv&=& -\frac{1}{4f_\pi^2}\left[\frac{s}{\sv^2}
\left(4\Delta J_2\stsv-S_0^2\Delta J_0\stsv-2\Delta
F_\beta\right)\right.\nonumber\\
&+&\left.(4m_\pi^2-s)\Delta J_0\stsv-4\Delta F_\beta\right]\nonumber \\
\Delta G_s^{(1)}\stsv&=&
\frac{S_0}{4f_\pi^2\sv^2}\left[\frac{3s}{\sv^2} \left(4\Delta
J_2\stsv-S_0^2\Delta J_0\stsv-2\Delta
F_\beta\right)\right.\nonumber\\
&+&\left.(4m_\pi^2-s)\Delta J_0\stsv-4\Delta F_\beta\right]
\label{finalff} \ea

As in \cite{glp02}, for a $T$-dependent quantity we denote $\Delta
H(T)=H(T)-H(0)$. Recall that all the $D\rightarrow 4$ UV
divergences are contained in the $T=0$ part
 \cite{gale84}
that is finite and scale independent  once expressed in terms of
the finite and scale independent constant $\bar l_6$ and the
subtracted $T=0$ loop integral $\bar J(s)=J_0(s)-J_0(0)$
\begin{equation}
F_t(s)=F_s(s)=1+\bar J(s) \frac{s-4m_\pi^2}{6f_\pi^2} +
\frac{(\bar l_6-1/3)s}{96\pi^2f_\pi^2}\ . \label{fft0}
\end{equation}

We remark that our finite temperature additions (\ref{finalff})
are written in Minkowski space-time for continuous $S_0\in\IR$. In
addition, we have used the on-shell condition $p^2=(p')^2$, valid
to one loop. As we mentioned in section \ref{sec:ftff}, the
one-loop form factors do not depend on $q_0$. For instance, there
are terms in $\langle V_0 \rangle$ proportional to
$\vec{q}\cdot\vec{S}=q_0 S_0$ to this order that would otherwise
give a nontrivial $q_0$ dependence. Finally, note that in the
above expression and to this order we can use either the physical
$f_\pi$, $m_\pi$, their tree level values or their $T$-dependent
ones, since the differences are of higher order. We have chosen to
write down our expressions in terms of the $T=0$ physical values
$f_\pi\simeq 92.4$ MeV, $m_\pi\simeq 139.6$ MeV.

It is easy to check the consistency of our explicit
one-loop form factors (\ref{finalff}) with the gauge identity
(\ref{ginlo}), showing how $G_s^{(1)}\neq 0$ for arbitrary
$S_0,\sv$. Another interesting check concerning the unitarity of
(\ref{finalff}) in the center of mass frame will be analyzed in
section \ref{sec:pu}.

\subsection{Pion electromagnetic charge and radius at $T\neq 0$.}
\label{charge}

A direct prediction of our ChPT calculation is the pion
electromagnetic static charge density  at $T\neq0$ and low
energies. Recall that at $T=0$, the total pion charge and charge
radius are $\langle Q\rangle_0=F(0)$ and $\langle r^2
\rangle_{0}=6F'(0)/F(0)$ where $F(s)$ is the form factor. Thus,
from
 (\ref{fft0}), $\langle Q\rangle_0=1$ and $\langle r^2 \rangle_0=(\bar l_6
-1)/16\pi^2f_{\pi}^2$. In fact, this is the simplest way to
estimate the value of $\bar l_6$ in ChPT to one loop \cite{gale84}
giving $\bar l_6=16.5\pm1.1$. Recent evaluations, including two
loop corrections \cite{bijnens}, obtain $\bar l_6=16.0\pm 0.5\pm
0.7$, where the last error is purely theoretical and dominates the
uncertainty. Concerning the charge radius, the latest experimental
average quoted in the PDG is $\langle r^2\rangle_0=0.45\pm 0.01$
\cite{pdg}.

Likewise, at $T\neq 0$ taking into account  the Lorentz covariance
breaking:

\be \langle r^2 \rangle_T=-\frac{6}{Q_T}\left.\frac{d
F_t(0,\sv)}{d \sv^2}\right\vert_{\sv=0} \ee where $\displaystyle
Q_T=\lim_{\sv\rightarrow 0^+}F_t(0,\sv;T)$. Note that the charge
density is defined through $V_0$ and hence $F_t$ in
(\ref{ffgenform}) must be used. We consider the spacelike radius
as would be measured in $t$-channel $e-\pi$ scattering and
therefore $S_0$ must be set to zero before $\sv$ (static limit).
Analogous timelike and magnetic moment radii could also be
defined. From equation (\ref{finalff}) we find: \ba
\label{changecharge} Q_T&=&1-\frac{1}{2\pi^2
f_\pi^2}\int_{m_\pi}^\infty dE
\frac{2E^2-m_\pi^2}{\sqrt{E^2-m_\pi^2}}
\ n_B(E;T)\nonumber\\
Q_T \langle r^2 \rangle_T&=&\langle r^2 \rangle_0+\frac{1}{12\pi^2
f_\pi^2}\int_{m_\pi}^\infty \frac{dE}{E^4
\sqrt{E^2-m_\pi^2}}\left[
3(E^2-m_\pi^2)(E^2+2m_\pi^2)\ n_B(E;T)\right.\nonumber\\
&+&\left. E(2E^4+2m_\pi^4-E^2m_\pi^2)\ \frac{d n_B(E;T)}{dE}
\right] \label{qrt}\ea where
$n_B(x;T)=\left[\exp(x/T)-1\right]^{-1}$ is the Bose-Einstein
distribution function.

\begin{figure}[h]
 \hspace*{2cm} \hbox{\psfig{file=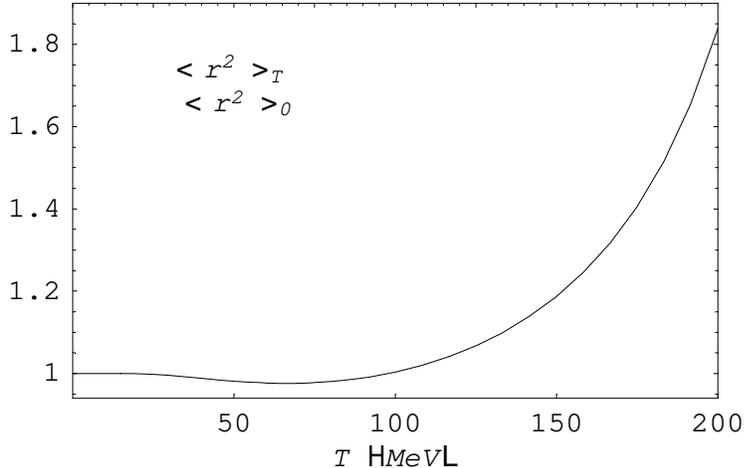,width=10cm}}
\caption{\rm \label{fig:radio} The electromagnetic pion charge
 radius at $T\neq 0$.}
\end{figure}

Note that the positive charge decreases with temperature due to
$\Od(T^2)$ corrections in the $s\rightarrow 0$ limit. This is
similar to the electric charge Debye screening  in QED
\cite{lebellac,kapusta}. For the negatively charged $\pi^-$, the
whole  eq.(\ref{changecharge}) changes sign. However, in absolute
value both charges decrease by the same amount and the gas remains
neutral. We have plotted  $\langle r^2 \rangle_T/\langle r^2
\rangle_0$ in Figure \ref{fig:radio}. The $T$-correction is almost
negligible below 100 MeV, where it decreases very slightly. For
higher $T$, the radius increases considerably, the dominant
contribution coming from the $Q_T$ screening discussed above.
Since they rely on ChPT alone, these results are model
independent. Our analysis confirms earlier results for the charge
radius based on extrapolated QCD sum rules \cite{dom94}.

We get a rough estimate of the deconfinement temperature $T_c$
\cite{kapusta} when the pion electromagnetic volume equals the
inverse pion  density,  i.e, $ (4\pi/3)\langle
r^2\rangle_T^{3/2}=1/n_\pi(T)$, where
 $\displaystyle n_\pi (T)=3\int \frac{d^3\vec{k}}{(2\pi)^3}
 n_B(\sqrt{\vert\vec{k}\vert^2+m_\pi^2};T)$. Taking just $\langle r^2\rangle_0$
  gives $T_c\simeq$ 265
 MeV, which is clearly too high, as commented in \cite{kapusta}.
 Thus, the thermal increase of $\langle r^2\rangle_T$
 reduces $T_c$ to a more realistic value. With our above result we get
 $T_c\simeq$ 200 MeV, which in fact is closer to the {\em chiral restoration} transition
 temperature as estimated within ChPT \cite{gelejr}. The uncertainties in $\bar l_6$ amount to
 $\pm$ 4 MeV.
 This critical temperature is below the
 temperature where $Q_T=0$ and the radius diverges, which would correspond to a vanishing
  pion thermal mass and is clearly beyond the validity range of the
   chiral expansion. As a matter of fact, these estimates have to be regarded as merely
    qualitative.

It has also been shown \cite{gale84,Res} that the value of $\bar
l_6$ is saturated by the $\rho$ resonance contribution, that,
together with VMD, implies
 $\bar l_6 \simeq 96 \pi^2  f_\pi^2/M_\rho^2$ \cite{Res}, up
to chiral logarithms.
Therefore,
  the electromagnetic radius behaves as $M_\rho^{-2}$. Thus, at
  $T\neq 0$, our previous results, still model-independent,
   would suggest an  almost constant (though slightly increasing)
   $M_\rho (T)$ for very low $T$ and a clearly decreasing $M_\rho (T)$ for $T\gsim$ 100
   MeV. This is indeed the behaviour that we found in  \cite{dglp02} and is also
   confirmed by our analysis in section \ref{uni}. Such behaviour
   for $M_\rho (T)$ at very low $T$ is  a prediction of chiral
   symmetry and resonance saturation, at least in the chiral limit
   \cite{dei90}.

\subsection{Thermal perturbative unitarity in the center of mass frame}
\label{sec:pu}

Let us reduce our previous expressions to the center of mass frame
(c.o.m.), i.e, $\vec{p}=-\vec{p'}$ in (\ref{ffgenform}). This
amounts to look only to back to back $e^+e^-$ pairs in the
dilepton spectrum. The dilepton rate is particularly simple in
that case \cite{gaka} although the three-momentum distribution may
be important when performing more realistic analysis \cite{sb}.

Therefore, we take the limit $\sv\rightarrow 0^+$ in our previous
expressions. Noting that $J_2=(S_0^2/4)J_0+F_\beta/2+\Od(\sv^2)$
\cite{glp02} we see in (\ref{finalff}) that $F_t^{(1)}$,
$F_s^{(1)}$ and $\sv^2 G_s^{(1)}$ have a finite c.o.m. limit,
which is reassuring. In addition, in the c.o.m. frame
$(p-p')(p+p')=0$ implies  $S_k q_0 G_s=S_k  (2S^j p^j/S_0) G_s=2
p_k(\sv^2 G_s/S_0)=q_k(\sv^2 G_s/S_0)$ so that $\sv^2 G_s/S_0$ can
be reabsorbed in $F_s$ in (\ref{ffgenform}). Finally, we find: \ba
\lim_{\sv\rightarrow 0^+} \Delta F_t^{(1)}\stsv&=&
\lim_{\sv\rightarrow 0^+}{\Delta
F_s^{(1)}\stsv}=\frac{1}{6f_\pi^2}\left[\left(S_0^2-4
m_\pi^2\right)\Delta J_0 (S_0,\vec{0})+4\Delta F_\beta\right]
\nonumber\\
\lim_{\sv\rightarrow 0^+} \sv^2 G_s^{(1)}\stsv&=&0 \label{ffcm}
\ea

 Therefore, in the c.o.m. frame there is only one $T$-dependent
form factor, which we will just call $F(S_0;T)$ for simplicity.
The tadpole contribution $F_\beta$ is real, while $J_0$ contains a
nonzero imaginary part required by unitarity. In fact, following
the same steps as in \cite{glp02}:

\be \im F^{(1)}(E+i\epsilon;T)=\sigma_T(E)
\frac{E^2-4m_\pi^2}{96\pi f_\pi^2}\label{pertunit}\ee with \be
\sigma_T (E)=\sigma(E^2)\left[1+2n_B(E/2;T)\right] \ee for
positive energies above the two-pion threshold $E>2m_\pi$. Here,
$\sigma(s)=\sqrt{1-4m_\pi^2/s}$ is the two-pion phase space
factor.

At $T=0$, (\ref{pertunit}) is the  perturbative version of the
form factor unitarity relation $\im F (s) =\sigma (s) F^* (s)
a^{11} (s)$ where $a^{11}$ is the $I=J=1$ partial wave projection
of the $\pi\pi$ scattering amplitude (to lowest order $F=1$ and
$a^{11} (s)=(s-4m_{\pi}^2)/96\pi f_\pi^2$). At $T\neq 0$, the
correction factor $1+2n_B=(1+n_B)(1+n_B)-n_B n_B$ is interpreted
as enhanced phase space \cite{weldon} due to the difference
between induced emission and absorption processes \cite{glp02}.
Therefore, we find that the  form factor satisfies
 a perturbative unitarity relation analogous to that in the $T=0$ case,
but now in
 terms of a thermal phase space factor. The same happened with the
  thermal amplitude in \cite{glp02}. This is not only a consistency
 check of our calculation, but it will be the basis of our
 unitarization method used in the next section in order to
generate the $\rho$ pole in the form factor.

\section{Unitarization and applications}
\label{uni}

The ChPT perturbative form factors analyzed in the previous
sections provide the prediction of  chiral symmetry  to
 next to leading order  at finite temperature in a model-independent
 way. However, by
construction, they cannot
 reproduce a pole or resonant behaviour, which in the
$I=J=1$ channel
 corresponds to the $\rho$. Our approach here will be to construct
 a nonperturbative thermal form factor $F$ imposing {\em exact}
 thermal unitarity with respect to the nonperturbative pion
scattering amplitude.
The latter is well approximated by the Inverse Amplitude
 Method (IAM) \cite{iam} derived at $T\neq 0$
 in \cite{dglp02} $a^{IAM}(E;T)=a_2^2(E^2)/[a_2(E^2)-a_4(E;T)]$ where
 $a_2$ and $a_4$ are the ChPT partial waves to tree and one-loop
 level respectively (we are suppressing the 11 superscript) calculated in
 \cite{glp02}. The IAM
amplitude is also exactly unitary, i.e,
 $\im a^{IAM}=\sigma_T\vert a^{IAM} \vert^2$ so that our only
 physical input, apart from chiral symmetry, will be unitarity. It
 should be borne in mind that our approach of demanding exact
 thermal elastic unitarity is meant to be valid for energies and
 temperatures such that $n_B(E/2;T)$ remains small \cite{dglp02}.
 For the  dilepton spectrum this means
 that for typical freeze-out temperatures $T\simeq$ 150 MeV,
  our approach is rather accurate around the $\rho$ scale, which is
precisely where unitarity is saturated. Near the pion pair
threshold, corrections to pion propagation not included in this
work (nominally of $\Od(p^6)$ in ChPT) have been conjectured to be
more important \cite{koso96,sb,gaka,soko96}.

\begin{figure}[h]
\hbox{\psfig{file=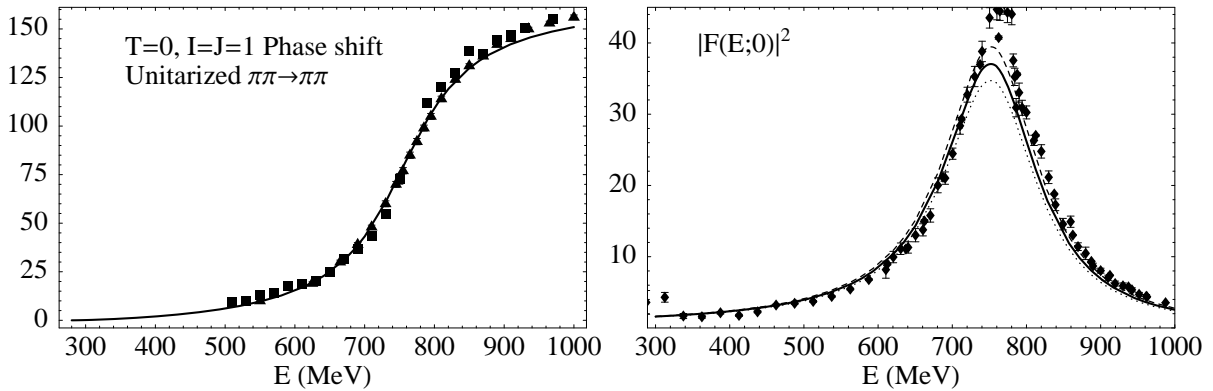,width=16.cm}}\caption{\rm
\label{fig:ff0} a) We show the phase of the $a^{IAM}$  for elastic
amplitude of $\pi\pi$ scattering in the $\rho$ channel. The data
comes from \cite{rhodata} b)  The unitarized form factor at
 temperatures  $T=0$, for $\bar l_6=18$ (solid line),
$\bar l_6=17$ (dotted line) and $\bar l_6=19$ (dashed line). The
data comes from \cite{Fdata}.} \end{figure}

If we take $F \propto a^{IAM}$ with a real proportionality
constant we readily guarantee the exact unitarity condition $\im
F=\sigma_T F^* a^{IAM}$ and also that both the amplitude and the
form factor have the same poles in the complex plane as well as
the same complex phase (phase shift), as it happens for $T=0$.
Imposing the correct low energy perturbative ChPT expansion
$F=1+F^{(1)}+\cdots$ fixes the real proportionality constant, so
that:
 \be
F(E;T)= \frac{1+\re
F^{(1)}(E;T)}{a_2(E^2)+\re
a_4(E;T)}\,
\frac{a_2^2(E^2)}{a_2(E^2)- a_4(E;T)}.
\label{ffunit}
 \ee
 valid for $E>2m_\pi$, where perturbative unitarity holds.
The above formula was developed at $T=0$ in the strongly
interacting electroweak spontaneous symmetry breaking sector
\cite{dhpr00}. At low energies it reproduces the chiral expansion
up to $F^{(1)}$ plus terms of higher order, which should be
smaller. Thus we expect that it should reproduce also the low
energy data with an $\bar l_6$ value slightly different, but
reasonably close, to that used in the previous section with the
pure one-loop ChPT. All that remains is to adjust the undetermined
low-energy constant $\bar l_6$ to zero-temperature experimental
data and the finite $T$ behaviour follows as a prediction. For
$a^{IAM}$ and hence for $a_2$ and $a_4$ we are using the very same
calculation given in \cite{glp02,dglp02} but note that in order to
obtain the unitarized $\vert F(E;T)\vert$ from (\ref{ffunit}) we
need the real part of the form factor, whose perturbative
calculation we have carried out  in section \ref{sec:loopcal}. In
Figure 3a we show how the $\rho$ data \cite{rhodata} are nicely
described by the resulting phase shift in the $11$ channel, common
to the amplitude and the form factor. This fixes the form factor
phase. In Figure 3b. we plot $\vert F(E;0) \vert^2$ for 0.3 GeV
$<E<$ 1 GeV compared with the data in \cite{Fdata}. The solid line
has been obtained with $\bar l_6=18$, that describes data
reasonably well and is compatible with the perturbative value
quoted in section \ref{charge}. For illustration we also provide
curves with $\bar l_6=17$ (dotted) and $\bar l_6=19$ (dashed).
Observe that the resonant behaviour around $M_\rho\simeq 770$ MeV
is clearly reproduced in both figures. In fact, note that the
unitarized formula gives exactly the form factor phase and
therefore the position and width of the $\rho$ pole is very
accurately reproduced. However, the modulus is more subject to
perturbative uncertainties, as reflected in Figure \ref{fig:ff0}.

\begin{figure}[h]
 \hspace*{2cm} \hbox{\psfig{file=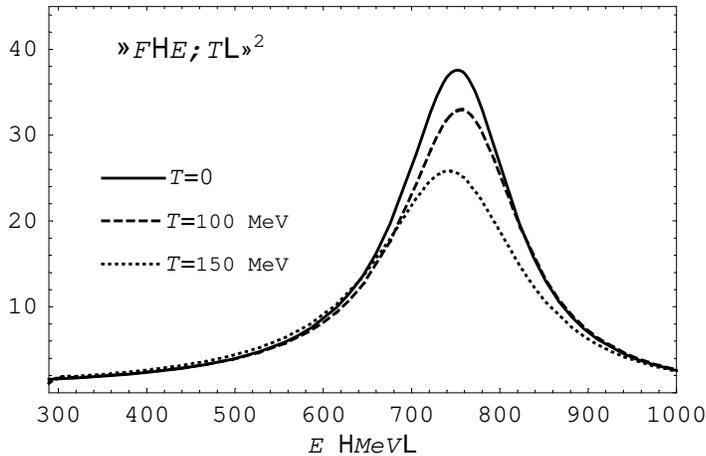,width=10cm}}
\caption{\rm \label{fig:fft} The unitarized form factor at
 temperatures  $T=0,100,150$ MeV.}
\end{figure}

Next, in Figure \ref{fig:fft} we have plotted $\vert F(E;T)
\vert^2$ for different temperatures. We observe that the form
factor decreases and widens with temperature. The mass position of
the peak  moves slightly to the right  and then drops
 in the $T=150$ MeV curve, consistently with our discussion
on the pion electromagnetic radius in section \ref{charge}. Note
that, by construction, the form factor (\ref{ffunit}) has a peak
exactly at the same place and with the same width as the
amplitude, that we had already studied in \cite{dglp02}, and where
the $\rho$ pole moved further away from the axis, explaining the
strong flattening of the form factor in this work.

Our results broadly agree with \cite{soko96}. Note however that
 we have not introduced
explicit resonances, the physical assumptions are just chiral
symmetry and unitarity. Our $T=0$ peak also falls a little bit
short of the data but is much closer than \cite{soko96}. Since we
only deal with two pions, and therefore only the $\rho$ resonance,
this could be partially due to $\omega$ contamination in the data
coming from $e^+e^-$ annihilation (first reference in
\cite{Fdata}). Indeed, the lowest data point at the peak, closer
to our curves, comes from $\tau$ decay (second reference in
\cite{Fdata}), where no $\omega$ can be produced. The two-loop
calculation \cite{hannah,bijnens} can also improve the situation,
but this is beyond our scope at $T\neq0$.

The modulus of the form factor enters directly in the dilepton
rate from pion annihilation. The simplest approach is to use
kinetic theory to leading order, so that the dilepton rate is
written as an integral containing  the pion distribution functions
and $\vert F\vert^2$ \cite{kaj86,gaka}. In the c.o.m. and in
thermal equilibrium the rate is simply proportional to
$n_B^2(E/2;T) \vert F(E;T)\vert^2$ \cite{gaka}. However, the
equilibrium result is not enough to  fit the experimental data.
The dynamics of the plasma expansion, the contribution of other
channels and the experimental acceptance have to be properly
accounted for.  Let us remark that, as pointed out in
\cite{koso96,soko96}, it might be crucial that there is a
quark-hadron phase transition and a rather long-lived mixed
 phase in order that medium modification effects
coming from pion annihilation are sizable in the dilepton yield.
In \cite{sb} it has been found that to fit the dilepton data
reasonably well,  medium effects need to be included in the form
factor. Otherwise, the theoretical prediction would exceed the
dilepton data around $M_\rho$ which means that medium effects
should decrease the height of the $\rho$ peak. Moreover, the form
factor is expected to spread by a factor of two at $T\simeq 150$
MeV \cite{sb} and the peak position to shift to lower energies, as
in earlier treatments \cite{li95}. More elaborated space-time
analysis including also baryon effects \cite{rawa99} conclude that
the $\rho$ spreading allows to explain the CERES data, without a
sizable mass decrease. All these effects  are qualitatively
visible in our theoretical result for the form factor, our peak
mass position shift being  smaller than assumed for instance in
\cite{sb}. In this respect, it is worth pointing out that a recent
measurement by the STAR collaboration \cite{star} of the $\rho^0$
in-medium properties directly in
 $\rho^0\rightarrow \pi^+\pi^-$ shows a  softer $\rho$-mass
 decrease (of about -40 MeV) and therefore much closer to the claims
 in \cite{rawa99} and also to the size
 of our predictions here.
This is an important measurement, as chiral symmetry restoration
requires that $\rho$ and $a_1$ become degenerate \cite{pisarski}
so that this predicts a sharp $a_1$ mass decrease. Further
lowering of the $\rho$ mass as measured in pion observables can be
achieved theoretically due to various minor medium effects
\cite{pratt} such as collisional broadening, the Boltzmann factor
and rescattering at later stages of the collision (lower
temperatures).

\section{Conclusions}

We have analyzed the pion vector form factors at finite temperature in
Chiral Perturbation Theory. The general structure of the form
factors at $T\neq 0$ allows in principle for three different form
factors. However, the gauge Ward-Takahashi identity constrains
them, relating the form factors to the in-medium pion
dispersion relation.

Our explicit one-loop calculation gives the  two different form
factors not tied by gauge invariance with the correct $T=0$ limit.
In the center of mass frame, the two form factors coincide and
satisfy a perturbative thermal unitarity relation in terms of a
thermal phase space, consistently with our previous results on the
thermal $\pi\pi$ scattering amplitude. Using only ChPT, we have
also studied how the effective charge and charge radius of the
pion change with temperature. The effective charge is screened
with $T$, while the radius is almost constant for low $T$ and then
increases. Up to here, these results rely only on thermal field
theory and ChPT and are therefore model independent. We have also
checked that this behaviour is consistent with the expected $\rho$
mass thermal behaviour. Our form factor has also allowed us to
obtain a naive estimate of the deconfinement temperature.

On a second stage,  by imposing exact thermal unitarity
while respecting the ChPT low energy expansion,
we construct a
nonperturbative thermal form factor that reproduces previous
theoretical analysis and whose behaviour is qualitatively compatible with the
 observed dilepton spectrum. At the typical freeze-out
 temperatures of $T\simeq$ 150 MeV, our result predicts a
 peak height decrease and a spread of the form factor around the $\rho$ region,
 as expected from dilepton data. In addition,
 the position of the peak is slightly shifted to lower mass. We have arrived to our result
 imposing only chiral symmetry and thermal unitarity as physical
 assumptions.

The unitarization method discussed here is limited to the center
of mass frame. For future work, it would be interesting to extend
these ideas including the  three-momentum dependence needed for
the dilepton analysis. According to our discussion above, this
would need to account simultaneously for the effect of different
form factors and the dispersion relation. Still, an analysis of
back-to-back lepton data by ongoing experimental collaborations
would allow comparison with these very simple and powerful
theoretical results. In addition, a more realistic study should
 also include the space-time evolution of the plasma and baryon density effects.
  Work is in
 progress along these directions.

\subsection*{Acknowledgments}
 Work supported from the Spanish CICYT projects,
FPA2000-0956,PB98-0782 and BFM2000-1326.

\bibliography{apssamp}

\end{document}